# The impact of epistemology on learning: A case study from introductory physics

Laura Lising[*] and Andrew Elby

*Department of Physics, University of Maryland, College Park, Maryland 20742*

We discuss a case study of the influence of epistemology on learning for a student in an introductory college physics course. An analysis of videotaped class work, written work, and interviews indicates that many of the student's difficulties were epistemological in nature. Our primary goal is to show instructors and curriculum developers that a student's epistemological stance – her ideas about knowledge and learning – can have a direct, causal influence on her learning of physics. This influence exists even when research-based curriculum materials provide implicit epistemological support. For this reason, curriculum materials and teaching techniques could become more effective by explicitly attending to students' epistemologies.

## I. INTRODUCTION

In the past 15 years, physics education researchers have identified student difficulties in learning a broad range of physics concepts. Curricula targeting these difficulties have produced dramatically improved conceptual understanding.[1] In recent years, the physics education research community also has begun to look at student attitudes, expectations, and epistemologies (ideas about knowledge and learning).[2,3,4] For instance, students may think of physics knowledge as disconnected facts and formulas, or as interconnected concepts (often expressible as formulas). Students may think of learning physics as absorbing information from authority or as building up their own ideas.[5] This discipline-specific epistemology research builds on extensive research on more generalized epistemology.[6]

The recent focus on epistemology in physics education stems in part from two motivating ideas: (i) Students' epistemologies may affect their science learning. In that case, attending to epistemology may help us explain the variations in student learning outcomes with research-based curricula, create more effective curricula, and become better physics instructors. (ii) Fostering productive attitudes and epistemologies is in itself an important instructional outcome that could serve the students well beyond the course in question.

Our study addresses the first of these ideas and builds on previous research on college and pre-college learners. Most previous research has looked at correlations between epistemological measures and learning outcomes, finding that specific clusters of epistemological beliefs correlate with academic outcomes such as grade point average[7] and mathematical text comprehension.[8] In the physical sciences, one study found that certain epistemological beliefs correlate with integrated conceptual understanding in middle school,[9] while another found a correlation with ninth-graders' ability to reason on applied tasks.[10] In college physics, May and Etkina found correlations between students' gains on standard conceptual measures and their epistemologies as inferred from weekly written reflections on their own learning.[11]

A few studies have gone beyond these correlations to look at the causal influence of epistemology on students' learning behavior. These studies, generally carried out by observing students in the process of learning, have attempted to describe not just whether, but how learning is affected by epistemology and related factors. An excellent example is Hogan's thorough study on eighth-graders in which she observed relationships between students' "personal frameworks for science learning" and their social and cognitive engagement patterns during group learning.[12] Ryder and Leach's study found some correlations between college students' ideas about the nature of scientific knowledge and their self-reported activities during investigative project work.[13] Millar et al. observed that, among 9 to 14-year-olds, students' interpretations of classroom inquiry tasks varied according to their perceptions of the aims of scientific investigation.[14] Taylor-Robertson found differences in cognitive strategies used by college students according to their expectations of the meaningfulness of laboratory work,[15] and Edmondson found correlations between students' reported learning strategies and their epistemological stances as derived from interviews.[16] Dweck's work with students of varied ages showed some dramatic differences in learning behavior in the classroom which depended on students' ideas about the nature of intelligence.[17] And Hammer's study on college students described how students' ideas about knowledge and learning in physics affected how they solved physics homework problems during think-aloud interviews.[2] Taken together, these studies suggest a causal link between epistemology and learning and also raise new questions and issues. One issue is the distinction between personal and public epistemologies. Public epistemology encompasses a student's ideas about the nature of knowledge and learning for society as a whole - or for a disciplinary community. Personal epistemology concerns a student's ideas about her own knowledge and learning. A student's public and personal epistemologies can differ significantly. For instance, a student may doubt the possibility of coherence in her own knowledge (personal epistemology), but may expect scientists to seek and find coherence (public epistemology). Some of the previous correlation studies have looked at only one of these aspects of epistemology, while others have not made this distinction. Of the three "personal frameworks for science learning" in Hogan's research, one aligns fairly closely with personal epistemology while another aligns with public epistemology. She found that personal epistemology was linked strongly to the students' behavior, while public epistemology showed almost no effect. Thus her results point toward personal epistemology as being much more relevant to learning. For that reason, we focus on the personal epistemology of our student subject. This paper builds more on the work of Hogan,[12] Hammer,[2] and May and Etkina,[11] which focused on personal epistemology, than on the other studies mentioned above,[3,4,7-10,13-16] which looked at public epistemology or a combination of personal and public epistemology and other attitude-related variables.

To build on this line of research, we have done an in-depth and naturalistic case study of a single student to distill and carefully describe the likely causal mechanisms. Of course, a case study cannot produce definitive, generalizable results about causality. But it can add depth and detail to the perceptive toolkit of the instructor and curriculum developer by exploring specific causal mechanisms that might explain the correlations, and it can generate specific hypotheses about causal mechanisms for later testing in controlled-intervention studies. The following hypothetical example illustrates this point. Suppose a correlation is found between how quickly people learn rock climbing skills and how many safe exposures to heights they experienced as children. A possible causal mechanism underlying this correlation might be that lack of safe exposure to heights as children leads to a fear of heights, which then leads to some learners making more cautious movements. Case studies of a few slow-learning novice rock climbers might shed light on this hypothesis. As they first attempt new moves, do they give clues to their fear of heights verbally or physiologically? Can we rule out other possible causes by watching their behavior in detail? If so, the next step toward establishing causal mechanism might be a controlled-intervention study, safely exposing children to heights, enrolling them in a rock climbing class 15 years later, and comparing their learning speed to a control group who received a different intervention as children (for example, reading about rock climbing). Our goal is to develop a plausible existence argument and descriptive analysis for one particular causal mechanism between epistemology and learning, a mechanism that we hope will be tested in future controlled-intervention experiments.

The various previous studies we have cited also vary in the extent to which they disentangled students' personal epistemologies from their expectations about what's rewarded in a particular course. It can be difficult to distinguish between what a student thinks is productive for her learning and what she perceives is required by the teacher or the curriculum. Yet these can be quite disparate at times. Hammer's work with one student illustrates an example where a student ruefully and self-consciously abandoned her productive learning strategies to survive in a memorization-focused physics course.[18] A 1999 study by Elby gave some insight into the magnitude of the epistemology/expectations gap.[19]

Yet another issue arising in previous studies is the context-sensitivity of students' epistemologies. Survey-based research on students' epistemologies has established differences in approaches according to discipline, motivating research that is discipline-specific (such as Ref. 3). However, studies that involved observations of learning behaviors and studies with multiple epistemological assessments also uncovered a sensitivity of epistemology to context within a given discipline. Hogan, for example, found that epistemologies assessed in interviews differed from the approaches students took in class. One might expect this difference between students' tacit ideas and their explicitly articulated ones, but Hogan's interview methods included elicitation of tacit ideas through scenario-posing.[12] Thus it has become clear that taking context-sensitivity into account when designing studies and analyzing data is crucial in understanding epistemology and learning.

In our study, we look at a student, "Jan" and study both her personal epistemology and her learning and describe how one affects the other. By analyzing both epistemology and learning from the same set of classroom data, we avoid many context-related interpretive challenges and also provide a description that is immediately relevant to classroom learning and instruction. We use a separate set of data from interviews for a supplementary analysis, carefully accounting for context-driven differences and factors that point to public epistemology, expectations, and other influences. From this analysis, we are able to describe direct, causal links that are likely to exist between Jan's epistemology and her learning in the classroom. Due to the difficulty of making and describing such an in-depth argument about causality, we will do so for only one facet of Jan's epistemology, although Jan certainly possesses a wide array of ideas about knowledge and learning. We will focus only on how Jan selected and used conceptual resources in her physics learning, and not on other facets of her epistemology such as whether she treated knowledge as static or evolving.

After discussing our methods in Sec. II, we present in Sec. III two examples of Jan's classroom behavior in group work. In Sec. IV we use these examples to argue that a component of Jan's epistemology, her perception of a "wall" between formal reasoning and everyday/intuitive reasoning, contributes to her troubles learning the material. We then use an independent data set from clinical interviews to argue that Jan's epistemology does include this "wall." Section IV also addresses alternative, non-epistemological explanations of Jan's classroom behavior. Although some of those factors contribute to Jan's actions, we argue that no combination of them adequately accounts for her behavior, unless our epistemological explanation is included. This strengthens our case for a causal link between Jan's epistemology and her learning. In Sec. V we summarize this argument and discuss implications for instruction and research.

## II. METHODS

### IIA. Selection of our case study subject and collection of data

The subject of this case study, Jan, was a third-year student in the second semester of an algebra-based introductory physics course at the University of Maryland. The course, taken by about 100 students and taught by a physics education researcher, consisted of 3 hours per week of interactive lectures (including interactive lecture demonstrations[20] and other physics education research inspired elements), one hour of tutorial (worksheet-led conceptual group work[21]), and two hours of traditional-style laboratories. Jan had taken this course's prerequisite in the previous semester in a large lecture, purely traditional format from a different professor. Although we will highlight some of Jan's difficulties, overall she was a capable student. She has excellent mathematical skills, did well on the more traditional homework problems, and put in considerable effort, seeking help from peers. Some concepts she learned quite deeply while others she did not.

Jan was in one of the two groups of students we videotaped working in tutorials and laboratories over the course of the semester. For Jan's group, we had two usable hours of videotape. The other videotapes of her group were unusable because they were inaudible or because the discussion focused primarily on logistics rather than physics concepts. From among the students in her group, we chose to study Jan because she was neither a top nor a low-performing student, and because we believed that we were seeing epistemological indications in her behavior that we could explore with further analysis. (Again, since we are trying to make an existence argument and a descriptive analysis for a certain mechanism, rather than a generalizable conclusion, a random representative sample isn't necessary.) The following semester, she agreed to undergo six interviews about student reasoning with one of us (AE), whom she had not met previously. Over the following year the interviews were audiotaped and transcribed. Jan received $10 per interview. The first four interviews consisted primarily of Jan reasoning aloud in response to physics questions about real-world objects and phenomena. The final two interviews consisted of more formal, quantitative problems and of increasingly direct probes of Jan's epistemology.

## IIB. Analysis of the data and interpretation of the results

We reviewed the two usable hours of videotaped classroom data and looked for instances in which epistemology seemed to affect Jan's approach to learning and doing physics. From this review we developed a hypothesis about Jan's epistemology and its causal relationship to her learning. To test this hypothesis, we attempted to explain her classroom behavior in non-epistemological terms, by focusing on expectations (her perceptions of what is rewarded in the course), confidence, skills and habits, and the social dynamics in her tutorial group and in the interviews. We also used Jan's written homework to test predictions of the hypothesis we generated from the classroom and interview data. To quantify patterns in Jan's reasoning during the interviews, we developed and applied a coding scheme designed to pinpoint when she used formal, classroom-taught reasoning versus "everyday" and intuitive informal reasoning; when she was sense-making versus just trying to remember or throwing out ideas with little thought; and when she attempted to reconcile different lines of reasoning. We describe this scheme more fully in Sec. IV.

## III. CLASSROOM DATA

### IIIA. Episode 1: The electric field tutorial

The earliest usable video segment shows the students working for an hour on a tutorial developed at the University of Maryland to address student difficulties with the concept of the electric field. At this time in the semester, the four students (Jan, Veronica, Carl, and Nancy) have been working together for just a few weeks. Early in the tutorial, students find the electric force, F, exerted by a single, stationary source charge, Q, on several test charges, q, of differing magnitudes placed at the same point. They then work out the ratio of the force to the test charge, define the field, E, as the ratio, and then continue to explore which factors affect the field and which do not. The main point of this part of tutorial is that E expresses the influence of the source charge in a way that doesn't depend on the test charge used to "measure" that influence. The full transcript appears in the Appendix[22] of the electronically archived version of this paper.

Jan participates quite a bit, as do Veronica and Nancy. During the first part of the hour, Jan answers the tutorial worksheet questions using mathematical reasoning. Specifically, she reasons using the functional dependences between force, charge, and field in the relations $F = kQq/r^2$ and $E = F/q$. While doing so, she makes a series of errors that her group members and the teaching assistants catch and help her correct. After the students look at the forces, the worksheet asks them to describe the dependence of the electric field on the magnitude of the test charge. Veronica figures out that the field is independent of the test charge and Jan agrees with her. However, a few minutes later, Jan claims she doesn't get it, and explains her math reasoning. Veronica helps her and Jan eventually seems to understand.

> Veronica: It's the same ratio, cause the higher the test charge the bigger the force.
> Jan: Right, so they're proportional.
> [Veronica and Nancy digress for a while and then Veronica explains the ratio idea to Nancy.]
> Jan: I don't really get that, though. Cause like you know how you were saying that $E = F/q$. Cause like they're saying that that's -
> Veronica: It's force per test charge. So if you have a big test charge it's…
> Jan: I thought that meant that the electric field is gonna get, if you have a small one, then the E-field is gonna be big. But then if you have, cause you know, cause like my understanding is that it says like describes the ratio of the force felt by the test charge and the strength of the test charge, right?
> Veronica: Yeah the q changes, but that makes the force different.
> Jan: So it's not the E-field that changes but the force that changes.

When asked to consider the $E$-field at a different distance, Jan claims it cannot change. This time Nancy corrects Jan's math error.

> Jan: No, but what I am saying is E is equal to F over q, right? That doesn't include radius in it.
> Nancy: But F includes, um, includes r. Jan: Because further away is smaller.

When asked to consider how the field changes when the source charge changes, again Jan returns to $E = F/q$, this time reasoning as if $q$ represented the source charge rather than the test charge. Nancy tries to make a non-mathematical argument, but Jan ignores her.

> Jan: If q is on the bottom.
> Nancy: The point charge becoming smaller is the same thing as the distance becoming greater. It affects the outcome of this yet, so it's the same thing.
> Jan: If this [q] becomes smaller then that [F] becomes bigger. That's all it is, right?

Later, when the TA asserts that the field doesn't depend on the test charge, Jan protests that it does and Veronica agrees with Jan, using the erroneous math reasoning that Jan has been persistently using (reasoning with the formula $E = F/q$ while ignoring the functional dependence of $F$ on $q$), and Jan verifies that this is what she is thinking. Veronica immediately catches her own error, but Jan does not comment, continuing to appear confused.

It is important to notice that Jan is using mathematical reasoning that is sophisticated for this population. Confronted with an equation, she does not try to plug and chug. Instead, she tries to extract information from the functional dependencies of different quantities; she's good at attending to proportionalities and inverse proportionalities. However, despite her facility with mathematics, she makes several math errors. Although she is corrected each time and acknowledges her mistake, four times in a row, Jan makes the same math error, repeatedly reasoning with the equation $E = F/q$ without considering the fact that the force $F$ depends on distance and on the test charge, the very quantities she is being asked to vary.

Jans problem here seems to be her failure to check her mathematical reasoning against her common sense reasoning. Specifically, she does not link her math to a sense of physical mechanism in the way that Veronica does. For instance, it is highly unlikely that Jan would find sensible her prediction that the field does not change as the distance changes, were she to think it through intuitively. It is unlikely that she would continue to ignore the functional dependencies of $F$ in the relation $E = F/q$ were she attending to more than just mathematical accuracy each time a group member corrects her. In the interviews, Jan considers it obvious that greater distance leads to weaker fields. She momentarily acknowledges her understanding of this common sense idea in class by emphasizing "Because further away is smaller" in response to Nancy's comment. But then she ignores this idea, returning to reasoning with $E = F/q$ in isolation, as if the common sense ideas were irrelevant. Jan seems almost not to even hear the next attempt by Nancy to make a common sense argument, and then says about the mathematical reasoning, "That's all it is, right?"

We think this behavior is both a window into Jan's epistemology and evidence of how it is affecting her learning. In the following discussion, we propose that Jan's epistemology is causing her to act as if a "wall" separates formal reasoning from informal, common-sense reasoning and that this wall accounts for her lack of checking her mathematical answers against her informal understanding.

**IIB. The light and shadow tutorial**

Eight weeks later, the same four students are working on the Light and Shadow Tutorial.[21] The group has several light bulbs, a board with several small apertures, and a large screen. They manipulate the bulbs and apertures and observe the changes in the pattern of light on the screen. As the students work through the activities, the worksheet guides them to build a model of light to explain their observations. Early in the hour, the students observe that moving the bulb to the right causes the bright spot on the screen to move to the left, the opposite direction.

While attempting to answer the worksheet question, "What do your observations suggest about the path taken by light from the bulb to the screen?," Jan initiates a discussion of physical mechanism.

> Jan: So does that mean that the path is not a straight line? . . . Does that mean it's reflecting?
> Nancy: Oh, that's a good point. I don't know.
> Veronica: No, there's no mirror for it to reflect off.
> Jan: But it's not direct, right? Cause if it were direct, then wouldn't it move up when we move the thing up?
> Veronica: No. Because it's going like this. When you move it up. It's going through a hole. Well, I mean, I guess it is reflected light. Cause look, here's the hole. It's down here. It's going to go through the hole like that. You know what I mean? Because it's its position relative to that hole.
> Jan: But I mean, if it, if if it was direct, right, then the light wouldn't come through if it wasn't aligned.
> Veronica: If it was direct, then it would go like this. [gesturing horizontally with hand from light bulb, bangs into board above aperture to show that light would not pass through]
> Jan: Right. That's what I mean.
> Veronica: It would hit, it would just hit the board. Well, the light goes out, like that, like that. So, it's going to go, whatever path of that light it's going to hit right through the circle, it's going to keep going straight that way.
> Jan: Right.

Veronica seems confused about the "reflecting" and "direct" issues, but may realize that she and Jan might mean something different by "direct." So Veronica then goes on to explain in everyday language what she thinks is happening. Jan indicates that she understands. The dialogue continues with Veronica helping the group to understand how the model explains their observations.

> Nancy: How is it possible for the things to, like when we have the two bulbs, for one little circle to create the two . . .
> Veronica: Because they are two different directions. One's going in like that and one is going in like that.
> Jan: So you are saying that . . .
> Nancy: But what's the normal direction of the light? Cause that's what I'm asking.
> Veronica: It, it spans out, and whatever part goes through that circle is the part we're going to see.
> Jan: [drawing as she talks] So the light is like that and these are the rays, and the vector points that way will go through the hole.
> Nancy: Okay, so then if you move it up, then it's going to be?
> Carl: So if here is the hole and the light is down here, the light is going to go in the direction …
> Jan: Right, so like it has
> [Nancy, Jan, and Carl talking, unintelligible]
> Veronica: Really, it's just normal.
> Jan: All the rays are going like this. So, it's kind of like polarized.
> Veronica: Mmm, not really.

Jan's behavior here is puzzling. She is engaged and indicates her understanding of others' explanations, but she is using more technical and mathematical language: "rays," "vector," and "polarized." Her use of this last term is particularly striking. What is Jan doing in this exchange? Is she, like the others, trying to make sense of her observations and her group's explanations, or is something else going on? Veronica takes on this issue.

> Veronica: It's just, well, it's just, guys you're making it, you're trying to make it more difficult. It's just, the light goes out. It only goes through that one circle. So, obviously, if it is down here, and I'm looking through that circle. Look, you're sitting down here. You're looking at this big cardboard. You're looking through that little circle. All you're going to see is what's up there. It's a direct line.

In accusing the group of "trying to make it more difficult" than it really is, Veronica suggests that something other than simple sense-making is going on. Why would the group – and Jan in particular – do this? Fortunately, Jan tells us what she is up to.

> Jan: Look, I see what you're saying, alright. But, I'm just trying to make it like physics- physics-oriented. [laughs]
> Veronica: It is, it is physics-oriented. That's just the way it is.
> Jan: Alright.

Jan's behavior during this episode seems puzzling at first, but Jan is quite explicit in describing her motives. "I'm just trying to make it like, physics- physics-oriented." Her words and her behavior reveal her epistemology and its impact on her choices. Although she desists when challenged by Veronica, Jan strongly implies that she is not looking for an informal, common-sense explanation.

We should notice that, in searching for the more formal explanation and rejecting the common-sense one, Jan still claims to understand what the group has been discussing. ("I see what you're saying.") It may be that Jan is considering the intuitive explanation but searching for more technical language for the worksheet. Alternatively, it may be that Jan's rejection includes passing up an opportunity to understand intuitively. In isolation, her behavior here cannot distinguish these two possibilities. However, her homework sheds light on this issue by revealing a lack of understanding on Jan's part. The assignment asks students to apply the model for light they just constructed. When asked to predict the shape and size of a shadow, Jan draws straight lines indicating the rays of light from the bulb. However, her rays reach only the blocking object and do not extend all the way to the screen. She does not attempt to answer the question further and when asked to explain her prediction writes, "I don't know how else to think about it except for the rays from the light bulb." This directly contradicts the understanding she claims to have had during the discussion. If she really understood what Veronica was saying, she would have another way of thinking about it, the common-sense way Veronica keeps describing. ("Look, you're sitting down here. You're looking at this big cardboard ... All you're going to see ...") Veronica, using this understanding, gives complete and correct qualitative and quantitative responses on the homework, whereas Jan's formal labeling of the phenomena has not given her enough real understanding for either task.

When we combine this information from her homework with the statements she makes, the epistemological implications become clearer. As in the first episode (with $E = F/q$), Jan behaves as if *common sense reasoning is a separate endeavor from formal (mathematical or technical) reasoning, and that she considers only the latter to be "physics-oriented."*

## IV. DISCUSSION

### IVA. Preliminary hypothesis: Jan's epistemology places a barrier between formal and everyday reasoning

From these two episodes we make our preliminary interpretation. Jan's learning behavior seems to be strongly affected by her epistemology. In particular, Jan's epistemology divides the reasoning that can be used to understand physical phenomena into two disparate categories: formal, technical reasoning and everyday, intuitive reasoning. Between these two types of reasoning is a barrier (a "wall," metaphorically) that keeps Jan from looking for connections between ideas from the different sides.

Jan is quite adept at some types of formal reasoning. An open question for us at this point is how adept is she at informal reasoning. However, we claim that even if Jan were to use informal reasoning or accept that of her peers, her tendency not to link the formal and informal would continue to cause her difficulties.

In the following, we evaluate and expand our hypothesis in two ways: by analyzing additional interview data, and by exploring several counter-hypotheses (alternate interpretations that don't involve this epistemological mechanism).

### IVB. Results from the interviews: Jan uses everyday reasoning more often but still shows evidence of the epistemological barrier between formal and everyday reasoning.

Given Jan's behavior in the tutorials, what stands out most in the interviews is her willingness to approach problems using the kind of everyday knowledge and intuitive reasoning that we see her rejecting in tutorial. Often she immediately responds to a question using everyday/intuitive reasoning. Other times, if a formal line of reasoning doesn't work for her, she switches to everyday/intuitive reasoning. In the following example, from the third interview (3:243, interview 3, line 243), Jan is asked whether it matters if you choose a long wrench or a short wrench when loosening a stuck bolt.

> Jan: Well I think it matters, I definitely think it matters. Because one of the things that we did in physics was torque, and al-, you know when you have to draw like a lever arm? And um, I think it was $T = l\ r$, is that what it was? I don't know.
> Interviewer: What's $l$ and what's $r$?
> Jan: Like, like, okay this is like the pivot point, you know like here, and so you would draw like a line, and this is like the place from which you are going to change it, you know, or like, you know…
> Interviewer: This is like where you're holding…
> Jan: And you draw like some line here. I can't remember exactly. I should've learned physics better. I should keep these things in mind. So I think like the further you go out, you know, the easier, that's not to say you go all the way out, but as things, it's better if you have it here than if you have it here. I think what I can think about is like a door. You've got like the hinge here and you know you've got like the swinging door. I think if you push here [closer to the hinge] is the door is going to feel more heavy that if you push it out here [farther from the hinge].

Jan starts out here using formal reasoning, trying to apply the concept of torque and the (incorrect) formula $T = lr$. However, she doesn't seem to understand the formula or remember how to use it. So then, after implying that she doesn't trust her formal physics reasoning in this case, she switches gears and tries her common sense instead. Specifically, she comes up with an everyday experience (pushing a swinging door) that helps her solve the problem.

In contrast to her behavior in the tutorial, we find that Jan is far more likely to use everyday/intuitive reasoning to solve physics problems in interviews. This does not mean that Jan's preference for formal versus everyday/intuitive reasoning stems entirely from non-epistemological origins. Rather her epistemology might depend on context, a context dependence we can account for in a "resource-based" (as opposed to "belief-based") cognitive framework.[23] In the interviews, Jan may (consciously or unconsciously) activate an epistemological resource that guides her to use everyday/intuitive reasoning, a resource that stays "off" in the classroom context.

This context-sensitivity of Jan's epistemology is the subject of a separate paper.[24] We believe that this context-sensitivity is good evidence against the notion of epistemology as constructed of consistent, unitary, and context-insensitive beliefs, but we will not argue that point in this paper. Our goal here is to make a detailed plausibility argument for the impact of epistemology on learning. That argument relies on the detailed refutation of counterarguments. Thus we need to deeply analyze an aspect of Jan's epistemology that is fairly consistent across the two contexts from which our data is based. Although we do not take a "belief" approach, we will spend most of the remaining discussion describing an aspect of Jan's epistemology that is somewhat "belief-like" in these two contexts.

Our data show that Jan's view of the "separateness" of formal and everyday/intuitive reasoning is consistent across contexts. To get at this issue in the interviews, we look at the frequency with which Jan checks and reconciles multiple lines of reasoning. In the wrench example, for instance, Jan had an opportunity to reconcile her idea that torque is relevant with her idea that the wrench problem resembles pushing a door. As teachers, we want Jan to ask herself, "Does torque and the equation I thought of have anything to do with the door example I've given?" But Jan does not do that. We claim that her failure to do so stems in part from the barrier she places between formal and everyday/intuitive reasoning; if those two kinds of reasoning aren't connected, it makes no sense to try to reconcile them.

In contrast, Jan often reconciles two lines of reasoning when they're both formal or when they're both everyday/intuitive, For example, consider this exchange about a bowling ball swinging on a chain (1:33).

> Interviewer: Imagine that hanging from the ceiling by a chain is a bowling ball, and somebody gets it swinging back and forth, like a pendulum. And you've got like a stick or a mallet and you're allowed to whack as hard as you want, five times. And the purpose of you whacking it is to get it swinging as high as possible, so my question is, how--what would your strategy be for the whacking? How, and where in its swing would you want to whack it?
> Jan: I think I'd probably want to whack it when it's kind of like on its way up, and whack it like from the side going up, do you know what I mean?
> Interviewer: Can you draw me a little--?
> Jan: Yeah, like the chain is this way, and so it's on its way up, right? [Drawing a picture of a pendulum with the bowling ball at the lowest point.] So I'd probably whack it with the mallet right here.
> Interviewer: I see, and what's the, what sort of triggered you to think that?
> Jan: Well, it's already on its way up, so there's already force there, right? And if you just add force in the same direction, then it's probably just going to add up
> Interviewer: So it's like your adding onto something which is already there, as opposed to--?
> Jan: Trying to like oppose it, or to do something else.
> Interviewer: Right, so by that reasoning the very worst thing that you could do is like hit it the opposite way, going to, trying to beat it into going the other way.
> Jan: Right, right. I mean, this would probably like take a lot of energy out of you, but I think it would be good. [Laughing]
> Interviewer: Right, so each time you hit it, it would start going a little bit higher, and to that effect. Cool, OK, as long as we're on this big bowling ball pendulum, so let's say that you are done whacking it, now it's swinging higher, and--
> Jan: It's also like when you, have a person on a swing, well actually but when the person is on the swing you actually hit them when they're up here, but you push them that way, so… [Drawing a picture of a swing at it's highest point.]
> Interviewer: Huh, what do you make of that? [Laughing.]
> Jan: I don't know [Laughing.] well, I mean it's a little different when a person is on a swing because it's hard to get underneath them when they're like in this position.

Jan starts by reasoning that you should whack the ball when it's going fastest (at the lowest point in the swing) because you want to add "force" to "force." We code this as intuitive rather than formal reasoning; although she says the word "force," she seems to be using it in an informal, colloquial way, as a term that expresses the "energy" or "motion" of an object. We think she is reasoning, intuitively, that you want to add whatever you're going to add whenever there's the most of the target "stuff" already there.

Jan's second line of reasoning comes from her everyday experience. She thinks of pushing a person on a swing as an everyday instantiation of the problem at hand. But then she notices a conflict. When you push a person on a swing, you push at the high point of the swing, not the low point she has decided would be best for the bowling ball. Notice that rather than ignore the conflict by abandoning one or the other line of reasoning, Jan reconciles. She notes that, although we push people at the high point, this may not indicate the most efficient approach; we can't push the swing at its low point "because it's hard to get underneath them when they're like in this position."

Jan's successful reconciliation in the bowling ball example but lack of reconciliation in the wrench example is consistent with our epistemological hypothesis. She can and will reconcile when she doesn't have to overcome the barrier between formal and everyday/intuitive knowledge. This pattern emerged robustly in the coding results.

Thirty-six problems from the first five interviews were coded. (Interview 6 had a different format.) Each line of reasoning in each problem was coded as involving everyday/intuitive reasoning or formal reasoning. In the bowling ball example, for instance, we coded two lines of everyday/intuitive reasoning: the "adding 'force' to 'force' argument" and the "analogy to person on a swing." Altogether we coded 106 lines of reasoning. Jan used everyday/intuitive reasoning three times as often as formal reasoning, 71 versus 22. (We discuss the remaining 13 instances in the following.)

We then coded when Jan did or didn't reconcile given an opportunity (generally a conflict or lack of connection between two lines of reasoning.) Reconciliation opportunities involving two formal lines of reasoning or two everyday/intuitive lines of reasoning were coded as "within-type." For instance, in the swinging bowling ball example, Jan reconciles within-type between "adding 'force' to 'force'" and "analogy to person on a swing." Reconciliation opportunities involving formal versus everyday/intuitive lines of reasoning were coded as "between-type," as illustrated in the wrench example with the "torque equation" versus the "analogy to pushing a door." Of the 36 coded reconciliation opportunities, Jan reconciles about 40% of the time (14 reconciles.) To test our hypothesis about Jan's epistemology, however, we must look for differences between her tendency to reconcile within type versus between types.

Within type (28 coded opportunities), Jan reconciled about half the time (13 reconciles.)[25] Most are unprompted, and the others involve only mild prompting, as in the bowling ball problem when the interviewer says, "What do you make of that?" By contrast, between types (8 codings), Jan reconciled only once.[26] Although the number of codings involved does not give very reliable statistics, the dramatic difference between her within-type and between-type reconciliation rates, 46% versus 13%, supports our attribution of a barrier between formal and everyday/intuitive thinking in Jan's epistemology. A summary of our findings is included in Figure 1.

At times Jan's reasoning seems to be neither everyday/intuitive nor formal but rather a hybrid of the two. Thirteen of our 106 lines of reasoning were coded as hybrid. Hybrid reasoning is not a simple mix of the two other types. Rather, it is a type of reasoning in which the everyday/intuitive and formal are already integrated. (By contrast, between-type reconciliation is an action to address conflicts between everyday/intuitive and formal ideas that are not already integrated.) For the hybrid reasoning to occur, the ideas must have been integrated somehow *at some point in the past*, probably involving multiple between-type reconciliations. This type is rare in beginning physics students but is common in practicing scientists. The existence of a "hybrid" reasoning type seems to contradict our hypothesis that Jan views formal and everyday/intuitive reasoning as unconnected. However, our hypothesis claims not that it is impossible for these types of reasoning to be integrated for Jan, but rather that her epistemology generally prevents her from searching for these connections on her own. We found that most of Jan's integrated formal and intuitive reasoning can be explained in a way that is consistent with our epistemological interpretation. The instances of hybrid reasoning we found were either instances where the connection between the intuitive and formal were exceedingly transparent (almost unavoidable) or instances where the connections between the intuitive and formal had been stressed strongly and repeatedly in the course. We believe that this effort by the course instructor on particular topics helped Jan scale the epistemological wall that normally would have prevented these connections.

**Within-type Reconciliation Opportunities**

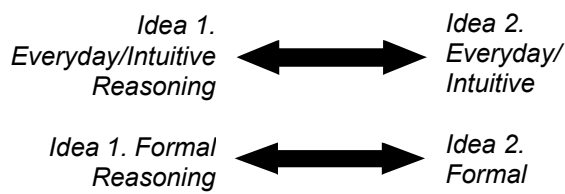

**Results: 13/28 opportunities reconciled (46%)**

**Between-type Reconciliation Opportunities**

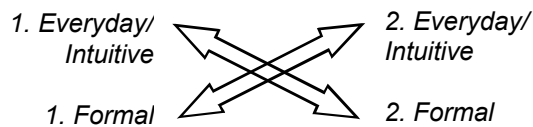

**Results: 1/8 opportunities reconciled (13%)**

**Figure 1.** Reconciliation opportunities occur when the relationship between two ideas Jan is using is ripe for examining. For instance, in the pendulum problem described above, the line of reasoning about adding force where the movement was greatest conflicted with the observation that you push a person on a swing at the turning points. If we call the adding force idea "Idea 1" and the swing idea "Idea 2," we see that in the diagram this is a reconciliation opportunity represented by the topmost arrow, since Idea 1 and Idea 2 both use everyday/intuitive reasoning. Since Jan does reconcile these two ideas, we count this in the results as a "within-type" reconcile. The differences in the relative rates of reconciliation in the within-type and between-type categories support our hypothesis of the wall Jan's epistemology places between the two reasoning types.

Our argument is that, due to Jan's epistemology, she does not search on her own for the connections between the two types of reasoning. As mentioned, in the interview coding, we found only one between-type reconciliation out of eight opportunities. This finding becomes more compelling and interesting if we look in detail at this one instance. Jan is asked to explain the motion of a pencil that has been thrown into the air (3:274). She starts with a line of reasoning that we coded as hybrid, integrating her everyday/intuitive ideas about a pencil's motion with formal concepts of velocity. Jan then discusses the influences of gravity and air resistance and the "force from your hand" in a hybrid fashion, but then gets confused about what happens to the force of the hand: Does it permanently die out at the peak or is it still influencing the motion when the pencil comes back down? At this point the interviewer pushes her very strongly not to let that go, but to try to figure out what happens to the force of the hand. Jan brings in her last, formal line of reasoning, discussing kinetic and potential energy, and uses it to explain how the "force that you added would be energy that you gave it, which is being interconverted in the system [between kinetic and potential]," and hence, the force (kinetic energy) from the hand is still there as the ball falls even though it died away temporarily at the peak (3:307). She is reconciling two lines of reasoning. However, this "between-type" reconciliation is weak in the sense that it was strongly prompted and involved a hybrid and a formal line of reasoning rather than fully scaling the barrier between formal and everyday/intuitive. Given that Jan's lone instance of a between-type reconciliation is weak, her tendency to reconcile only within type becomes a more robust coding result.

To test the reliability of our coding, we trained an education researcher not previously involved with the project and then had the coder apply our four coding steps to a subset of our data, some randomly chosen problems, and others deliberately culled from cases we ourselves considered particularly difficult. The external coder saw the same general trends as we did in the preponderance of everyday/intuitive over formal or hybrid reasoning and in the greater relative rate of within-type versus between-type reconciliation. To further validate this result, we then described the specific reconciliation opportunities from our original codings. For each case, the coder decided whether Jan reconciled, compromised, "checked," or didn't reconcile at all. The external coder's codings matched our codings 80% of the time.[27]

In summary, in comparison to the classroom data, the interviews yielded opposite results regarding Jan's inclination to use everyday/intuitive versus formal reasoning. The interviews also provide evidence that these two types of reasoning can be integrated by Jan under certain circumstances. However, both sets of data are robust in showing Jan's tendency not to look for connections between these two types of reasoning. What the interviews *cannot* establish completely is that this wall Jan places between everyday/intuitive and formal reasoning accounts for some of Jan's learning difficulties in the tutorials. To further that argument, we must refute alternative explanations.

**IVC. Refutation of alternative explanations: Jan's difficulties cannot be accounted for without epistemology.**

We now discuss other possible contributors to Jan's behavior in the tutorials. Although some of these factors play a role in Jan's actions, no combination of these effects, without epistemology, can account for all that we observe.

**i. Jan's difficulties are not due to lack of facility or confidence with mathematics.**

We might consider attributing Jan's repeated troubles using $E = F/q$ correctly in the electric field tutorial to a lack of facility or confidence with mathematics. However, evidence from that episode and from the interviews suggests otherwise. We have already noted that Jan reasons in a sophisticated manner with the equation $E = F/q$, using the functional dependencies between variables – in this case proportionalities and inverse proportionalities – to draw information from the equation.

Perhaps even more striking, Jan shows the ability to use mathematics intuitively. She refers to herself as "a proportions person" (2:194) and can formulate her own equations to express her intuitive reasoning (for example 1:128, 5:70), although she still differentiates these equations ontologically from the classroom equations. Furthermore, she explicitly states her confidence in her use and learning of mathematics (for example 6:144).

**ii. Jan's difficulties are not due to lack of skill with everyday reasoning in physics.**

One could argue that Jan simply lacks skill with informal reasoning in physics, perhaps for lack of practice. This might explain, for example, why she would fail to catch her math errors were she actually checking her math against her informal reasoning. However, weak informal reasoning cannot explain why she rejects as "not physics-oriented" the common-sense reasoning of the rest of the group. Furthermore, it is clearly not the case that Jan lacks this skill. In interviews, Jan proficiently uses common-sense, informal reasoning to work out physics problems.

**iii. Jan's difficulties are not due to lack of skill at checking.**

Perhaps Jan fails to check her interpretations of $E = F/q$, and fails to check how well she actually understands Veronica's model of light, because she has a general tendency not to check her reasoning, or because she's not good at checking. Again, the interviews suggest otherwise. In the 28 within-type reconciliation opportunities, there were also 6 instances we coded either as "checking only," when Jan makes sure one line of reasoning agrees with the *answer* yielded by another line of reasoning (without reconciling or even comparing the two lines of *reasoning*), or as "compromise," when Jan glosses over rather than directly addresses a contradiction she notices between two lines of reasoning. Our point is that, in 19 of 28 instances (68%), Jan notices and in some way addresses a connection between two lines of reasoning within type, proving that she often notices the potential tension between different lines of reasoning and that she has the skills to address the tension.

Jan's tendency to check and reconcile isn't confined to the interviews. For instance, later in the electric field tutorial (after the portions presented previously), the group considered a scenario in which a positive test charge is pushed directly toward a positive source charge. Does that "push force" do positive, negative, or zero work? Veronica reasoned that because "the potential energy becomes greater, the change in work is going to be negative," because the "The work … in, like, an electric field, it's the opposite, the opposite of the change in potential energy." Jan immediately wants to check this conclusion using an analogy the professor pointed out between electrostatic forces and gravity. She correctly notes that moving the point charge is analogous to changing the height of a mass, showing that the potential energy changes. In this instance in which she has two formal lines of reasoning (for example, reasoning about electric field and potential versus a professor-supplied analogy to gravitation), Jan wants to reconcile (although the group goes in a different direction before Jan can make progress).

**iv. Jan's actions cannot be fully explained by an expectation that only formal reasoning will be rewarded in the class.**

As previous studies show,[19,20] a student's expectations about what will be rewarded in a physics class need not align with her epistemological views about what constitutes learning and understanding. In tutorial, perhaps Jan focuses on formal reasoning to the exclusion of everyday/intuitive reasoning because she thinks formal reasoning is what the course requires. This counterargument also would explain why Jan is much more willing to use everyday/intuitive reasoning in the interviews, which were not part of the physics course. However, formal expectations cannot explain why Jan rejects everyday/intuitive reasoning that could help her use $E = F/q$ correctly. Even if exams reward formal reasoning only, they don't reward incorrect formal reasoning. And after being corrected, Jan acknowledges that she is using $E = F/q$ incorrectly. Her classmates suggest some common-sense reasoning that could help Jan apply $E = F/q$ correctly, but Jan acts as if that reasoning is irrelevant.

Similarly, in the light and shadow tutorial, "formal" expectations could explain why Jan rejects her group's everyday/intuitive model of light. But once again, expectations alone can't explain why Jan doesn't engage in learning just enough of Veronica-style everyday/intuitive reasoning to apply formal resources correctly. Formal expectations alone, unsupplemented by our epistemological interpretation, cannot explain Jan's behavior in tutorial.

**v. Jan's confidence in her informal reasoning does have an impact, but it can only account for her behavior in concert with her epistemology.**

Another counterargument is that Jan has low confidence in her ability to use everyday/intuitive reasoning in physics, and she therefore hides behind formal reasoning during group work. This lack of confidence could explain not only her nervous, perhaps self-deprecating laughter when she explains her quest for a more "physics-oriented" explanation, but also her willingness to use common-sense reasoning in the interviews, where she perhaps feels safer away from her peers and away from grade pressure.

We can quickly rule out one version of this counterargument, the idea that Jan hides behind formal reasoning during group work not because she lacks faith in her ability to use everyday/intuitive reasoning, but because she's afraid of "stepping out on a limb" by expressing her ideas publicly. According to this argument, Jan feels safer sticking to more objective, formal reasoning. This version of the counterargument fails, however, because it cannot explain why Jan rejects other students' everyday/intuitive reasoning in the light and shadow tutorial, or why she has such trouble using other students' qualitative ideas to help her interpret $E = F/q$ correctly.

Another version of the confidence counterargument has more traction: Jan avoids engaging in everyday/intuitive reasoning during group work largely because she lacks faith in her ability to learn and understand physics in those terms. This could help to explain her resistance to using everyday/intuitive reasoning when interpreting $E = F/q$ as well as her quest for a more "physics oriented" model of light. It would also explain why Jan seems so much bolder and more confident with her group when pursuing formal explanations such as an analogy between gravitational and electric potential. The interviews further buttress this counterargument. Perhaps because the interviewer repeatedly emphasized that he studies "student reasoning" and doesn't care whether her answers are correct, Jan willingly uses everyday/informal reasoning. Even then she often hedges her reasoning with qualifiers such as "maybe" and "it could be," and on several occasions she says her everyday way of thinking doesn't work reliably in physics. (We'll give an example in the following.)

It turns out that Jan's lack of confidence with everyday/intuitive reasoning in physics does indeed have a large impact on her behavior, as we have discussed. But it is Jan's epistemological stance that a barrier separates formal from everyday/intuitive reasoning that determines how she deals with her lack of confidence. She feels that her perceptions are imperfect and not to be relied upon.

> "It always seems like, you know, there's like a trick that I've missed, you know, something that I've overlooked or something that I haven't thought about." (5:138)

She gives static friction as an example, explaining that although you have ample experience with pulling things, you might fail to observe the need to pull harder at first (5:178). There are several ways a student might deal with the unreliability of perception-based everyday/intuitive reasoning. One response would be to use the formal and everyday/intuitive in conjunction, incorporating the two to make a more robust understanding, thereby avoiding the pitfalls of relying solely on imperfect intuitions and perceptions. But Jan does not take this stance. Instead, she generally keeps (unreliable) everyday/intuitive reasoning separate from (reliable) formal reasoning. Our point is that Jan's response to the unreliability of everyday/intuitive reasoning is driven in part by epistemology. Instead of seeing problematic everyday/intuitive reasoning as refinable and hence reconcilable with formal reasoning, she sees the two kinds of reasoning as too separate to inform each other.

At one point, Jan laments that physics is unlike chemistry, because chemistry is

> "kind of totally new, you know, like you kind of have a fresh brain,… I mean, you're talking about molecules and things you can't really see, you know so you have to kind of start fresh and I think so, so it makes it a little easier to think." (5:186).

Jan would rather start with a blank slate ("fresh brain") than try to refine and build on her own ideas. Jan's epistemology and confidence are entangled and mutually reinforce one another. Epistemology tells her the two domains are separable. Her confidence says to reject one, and then she only experiences success in the one domain, which leads to reduced confidence in the other domain, and so on. Although it is clear that confidence is playing a role, as an alternative explanation it cannot stand alone, because it relies on epistemology to have the effects we observe.

**IVD. Summary: The epistemological mechanism is plausible**

We believe that we have made a strong case that Jan's learning difficulties in the two tutorial episodes stem in part from her epistemology, in particular from the barrier Jan places between formal and informal reasoning. This barrier prevents her from searching for connections between these two types of knowledge. By examining supplementary data and evaluating alternative explanations,[28] we have established a highly plausible argument for this part of our hypothesis, that Jan's epistemology has a direct, causal effect on her learning.

In testing our hypothesis, we have also discovered more detail and subtlety to Jan's epistemology and its impact on her learning. We found that Jan's epistemology does not prevent her from using everyday/intuitive reasoning in some contexts (for example, interviews) and that there is a deep entanglement between Jan's epistemology and her distrust of her intuition and perceptions. Another intriguing issue is the existence in Jan's reasoning of a hybrid form in which formal and everyday/intuitive reasoning are integrated, generally when explicitly taught and sanctioned by the professor. Clearly this integration is not impossible for Jan. Yet her epistemology leads her only rarely to strive for this integration on her own, and even to resist it in many situations.

One hallmark of an epistemological effect is when we see students failing to use skills or knowledge they clearly possess. Jan's skills, her abilities, her store of ideas -- none of these are the "limiting reagent" for her learning in these episodes. She is capable and fluid with mathematical, technical, and everyday, common-sense reasoning. She is capable of checking her understanding and reconciling inconsistencies. She is capable of working through difficult problems for which she has very little relevant formal knowledge. Despite all these strengths, her epistemology sometimes gets in the way of her learning.

### IVE. Implications

Our case study has built on previous research into epistemology and learning to show, in causal detail, how epistemology can have a profound effect on the learning-relevant behavior of students. This is important for several reasons. First, this type of analysis provides supporting evidence that the effects of epistemology on learning outcomes observed in correlational studies are in fact causal. Learning also most likely plays a causal role in the development of a personal epistemology; but making a strong case for local causality in one direction is an important first step in understanding the complex interplay between the two.

For the curriculum developer and the classroom teacher, understanding how epistemology affects learning — or just keeping in mind that epistemology affects learning — has broad implications. In designing curriculum, developers must attend to epistemological as well as conceptual, social, or affective factors. For instance, epistemology can help us understand why a piece of curriculum optimized to address conceptual difficulties is ineffective for some students. Curriculum developers can take up the challenge of helping students associate their productive epistemological resources with the activity, the course, and the discipline.

There are also some implications that link specifically to the epistemological resource of a barrier between formal and everyday reasoning. Numerous physics teachers and researchers have noticed that students rarely hook up their conceptual/intuitive knowledge to their formal knowledge and problem-solving techniques. An epistemological barrier helps explain why this disconnect exists for many students, and what can be done about it.

For instance, Kanim found that even when students gain a deeper, more intuitive conceptual understanding of a topic (such as batteries and bulbs), they don't apply that knowledge to traditional quantitative problems (for example, about circuits).[29] To address this gap, Kanim started creating topic-by-topic *bridging worksheets* designed to help students connect their conceptual knowledge to their formal reasoning. Although many of these worksheets work well, Kanim noted the extreme, iterative effort needed to develop them; the bridging worksheets seem to keep exposing new student difficulties. For instance, even when students could answer an intuitive qualitative question about resultant vectors, they had trouble applying that knowledge to formal vector addition.

In our view, an epistemological barrier between everyday/intuitive and formal knowledge can help us understand why students have such trouble bridging those two types of knowledge, even when strongly supported.[30] If we are right, then Kanim's bridging worksheets might become even more effective by explicitly addressing this epistemological issue. The worksheets could include activities and reflection questions designed to help students realize that their thought processes sometimes incorporate such a barrier and that some of their "ah hah" moments of understanding occur when they scale or tear down the barrier. When students' epistemologies become more aligned with the goals of the bridging worksheets, students might become better at spotting and addressing new difficulties they encounter. For instance, consider a student who has learned to expect that her intuitive mathematical knowledge — for instance, her common-sense ideas about a hiker who walks two miles north and then three miles east — should mesh with formal mathematical tools. Especially when supported, that student would probably use her hiker knowledge when figuring out or making sense of formal vector addition rules. Consequently, she could resolve her vector difficulties more quickly than would otherwise be the case.

We lack space to discuss the details of epistemologically-focused curricula and teaching practices our research group has used.[31] Key components include listening for students' epistemological strengths and difficulties during office hours and recitation section; using tutorials and interactive lecture demonstrations designed to tap into productive epistemological resources (for sense-making and consistency) and highlight learning strategies; and using some reflective tutorial questions, homework questions, and class discussions to focus students' attention on their approach to learning.

### V. CONCLUSION

Our biggest challenge as instructors is listening to our students, responding to their difficulties, and facilitating their use of productive cognitive resources they possess. In diagnosing student learning, we must consider their strengths and difficulties of an epistemological nature. Specifically, we must learn to identify the epistemological resources that students possess and to understand which resources they are using during the learning process, so that we can help them to choose the more productive approaches to learning. Our strong argument about the plausibility of a causal mechanism by which epistemology can affect learning gives more reason than ever to believe that epistemological interventions could lead to better conceptual learning.

### ACKNOWLEDGEMENTS

We would like to thank Joe Redish, David Hammer, and the members of the Physics Education Research Group at the University of Maryland for playing a large part in this investigation, with special thanks to David May and Rebecca Lippmann. This work was done in part with the support of NSF grant REC-0087519. We would also like to thank advisory board members John Fredriksen and Paul Feltovich for useful suggestions early in the analysis. Thanks are also due to several anonymous reviewers and many members of the American Association of Physics Teachers whose questions helped us refine our focus.

---

.